# VARIATIONS OF PALEOINTENSITY IN THE PALEOGENE

**A. Yu. Kurazhkovskii[1], N. A. Kurazhkovskaya[2], B.I. Klain[3]**

[1, 2, 3]*Geophysical Observatory Borok, Schmidt Institute of Physics of the Earth of the Russian Academy of Sciences, Borok, Yaroslavl Region, 152742, Russian Federation*
[1]*E-mail:ksasha@borok.yar.ru*
[2]*E-mail:knady@borok.yar.ru*
[3]*E-mail:klain@borok.yar.ru*

**Abstract.** Behavior of the geomagnetic field intensity in the Paleogene from sedimentary rocks of the Russian Plate was investigated. It is revealed that in the beginning of the Paleogene alternating of paleointensity variations of little amplitude (about 0.5 $H_0$, where $H_0$ is the value of modern geomagnetic field taken as 40 µT) and bursts of large amplitude (up to 5 $H_0$) took place. At the end of the Paleogene paleointensity variations occur only with a little amplitude.

**Keywords:** Paleogene, paleointensity, sediments.

## 1. Introduction

Until recent time the conceptions on the geomagnetic field intensity in the Paleogene were based mainly on the data obtained by thermomagnetized rocks. These data allowed getting of rather rough conception on changes of the geomagnetic field intensity in the Paleogene. It has been shown that in the beginning of Paleogene the paleointensity was on average slightly higher than at the end of this period. For instance, *Solodovnikov* (1998) suggested that at the boundary of the Early - Middle Eocene (about 50 Ma) the abrupt (twofold) decrease in paleointensity occurred. According to the more recent papers (*Valet*, 2003; *Biggin and Thomas*, 2003; *Heller et al.*, 2003; *Tauxe and Yamazaki*, 2007) paleointensity in the Paleogene decreased really, but this change was not abrupt and considerable.

The sedimentary rocks are the preferable substratum for study of the paleointensity behavior. Attempts of a detailed study of the paleointensity behavior in the Late Paleogene by sedimentary rocks (in the interval 34 - 23 Ma) were made in papers (*Tauxe and Hartl*, 1997; *Constable et al.*, 1998). The referred authors concluded that in the Late Paleogene the paleointensity variations with characteristic times of the order of a million (a few million years) took place. The paleointensity of the Early and Middle Paleogene by sedimentary rocks has not been studied yet.

This paper sets to study the paleointensity behavior of the Early - Late Paleogene by sedimentary rocks from the southern part of the Russian Plate.

**2. Analyzed data**

Paleointensity behavior in the Paleogene was investigated by two sedimentary layers. One layer situated north of the Big Caucasian Ridge (coordinates 43°22′ N and 43°41′ E) and was disclosed by geomapping borehole (hereinafter deposits or sediments of borehole). Another layer was situated in the south of the Saratov Oblast (coordinates 50°42′ N and 45°39′ E). The samples of latter sedimentary layer were collected from the natural outcrop on the bank of the Volga River (the Saratov region sediments).

**2.1.** The formation of sedimentary layer disclosed by geomapping borehole was took place during the period from the Maastrichtian (end of the Cretaceous) to the Chattian (end of the Paleogene). The borehole was 570 m deep. The thickness of the Paleogene sediments was 530 m. The magnetic and petromagnetic properties of this sedimentary layer were studied in order to draw the magnetostratigraphic scheme of the Pre-Caucasian deposits. Earlier these studies were used for drawing the regional magnetostratigraphic scheme of the southern regions of European part of Russian Federation (*Bogachkin*, 2004) and for specification of the Cenozoic part of the magnetostratigraphic scale of the Phanerozoic (*Molostovskii et al*., 2007). The present study continues the above mentioned examinations. The samples of the Paleogene were collected from 280 stratigraphic levels with levels of 1 - 2 m interval. The time of sediment accumulation between the two nearest levels averaged 100 - 200 thousand years. This estimate is rather conditional. It is unlikely that the sediments of the region situated near tectonically-active Caucasus area were accumulated evenly. However, the detail of the sampling is sufficient detailed to detect paleointensity variations with the characteristic times of a million years order. According to *Bogachkin* (2004) and *Molostovskii et al*. (2007) examined sediments had predominantly reverse polarity. Against the background of the above polarity, 18 intervals of the negative polarity were found. This suggests that the deposits of this borehole are appropriate for the reconstruction of geomagnetic events in the Paleogene in good fullness.

To determine the direction of the natural remanent magnetization (NRM) two samples from each of the level were selected. In addition, to determine the inter-layer paleomagnetic concentration of NRM directions fives samples were collected from every tenth level. The measurement of the magnetic parameters of these deposits NRM, saturation isothermal remanent magnetization (SIRM), magnetic susceptibility (K) and the magnetic susceptibility after heating at the temperature above $500°C$ (TK), and thermal cleaning NRM were determined in the Paleomagnetic Laboratory of Scientific-Research Institute of Geology of Saratov State University.

The sedimentary layer consists of marine, mainly grey-colored carbonate and terrigenic deposits. The results of separation of grains of magnetic minerals and thermomagnetic analysis suggested that terrigenic magnetite is the main carrier of the sediments magnetization (*Bogachkin*, 2004; *Molostovskii et al.*, 2007). According to *Guzhikov et al.* (2003) the following criteria indicate the orientational nature of the sediments' NRM and, hence, their primacy: the presence of detrital grains of ferromagnetic minerals, low values of the Königsberger factor (Q < 0.1) and low inter-layer concentration of distribution of the vectors of remanent magnetization directions (k = 5 - 15).

The Lutetian deposits had weak magnetization. This is why they were not used for magnetostratigraphic studies and for the reconstruction of the paleointensity. In the present study we omit the data on Lutetian paleointensity.

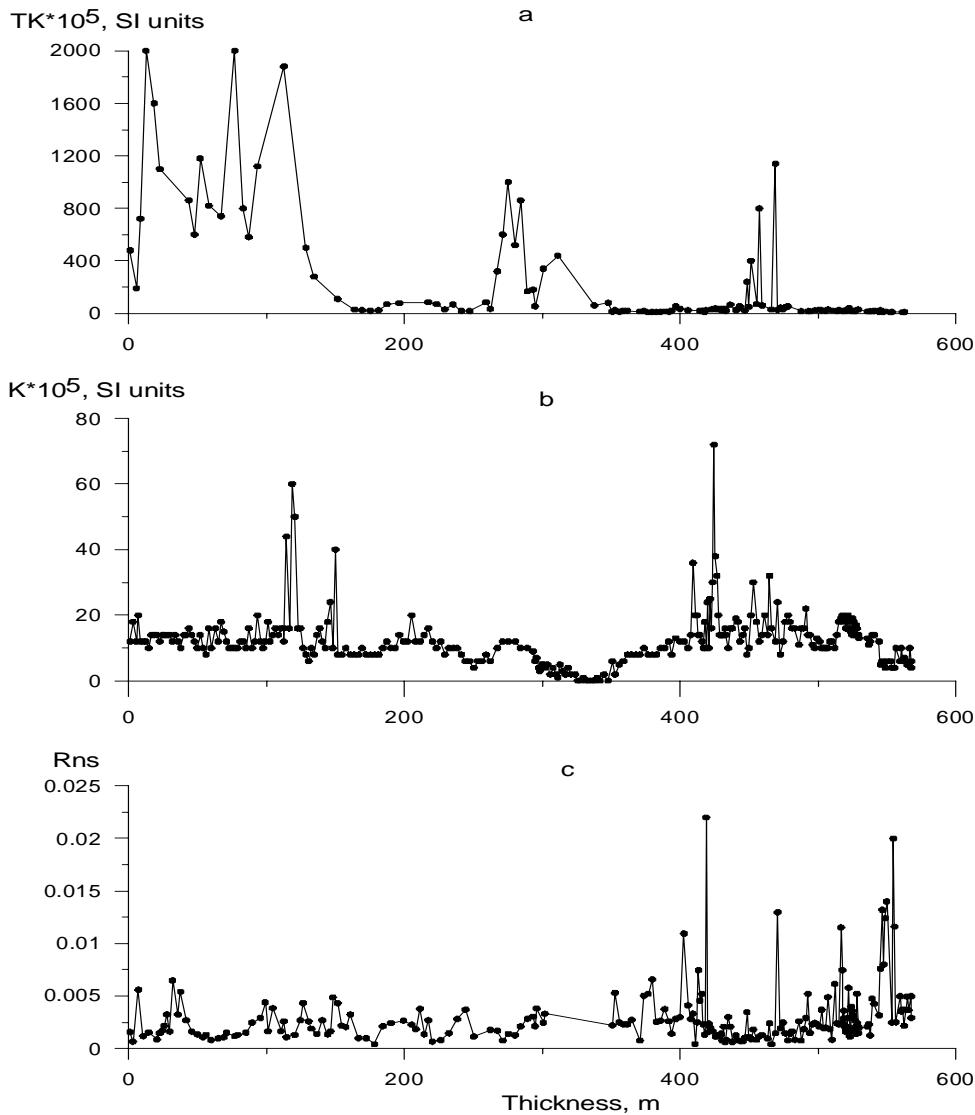

**Figure 1.** Variations in the petromagnetic (TK and K) and paleomagnetic (Rns) parameters in the Paleogene by sediments from geomapping borehole.

**2.2.** We identified the dynamics of paleointensity using the behavior of the parameter Rns = $NRM_t/SIRM_t$ (the parameters with index t have been measured after thermal cleaning). The changes of Rns, K and TK are shown in Fig. 1. As can be seen (Fig. 1), the behavior of Rns is not associated with changes of the petromagnetic parameters K and TK. This was yet another argument in favor of the suitability of the sedimentary layer to study of the paleointensity. It is worth noting that near the boundaries of geological epochs the magnetic susceptibility increased significantly. This indicate that at the boundaries of epochs not only the replacement of biota but and the replacement of geological processes took place. However, as it seen from Fig. 1 the replacement of these processes not was reflected explicitly on the behavior of the Rns parameter.

We utilized two ways for the calibration of Rns values or the determining of "absolute" values of paleointensity on deposits of borehole. In the first case, calibration of the behavior of the Paleogene paleointensity was performed using the results of the determination of paleointensity by thermomagnetized rocks. These data were taken from the database (DB) PINT10 [*http://earth.liv.ac.uk/pint/*]. The description of this database is given in (*Perrin and Schnepp*, 2004). Earlier this way the calibration of paleointensity obtained on sedimentary rocks was performed by *Valet et al.* (2005). For this purpose we used the ratio $\alpha = H_{DBmean}/Rns_{mean}$, where $H_{DBmean}$ is the average values of the paleointensity for geological age as calculated using of PINT10 data, $Rns_{mean}$ is the average values of the Rns for geological age. For the calculation of the paleointensity the following equation was used: $H/H_0 = \alpha*Rns/H_0$. In the second case, experimentally obtained relation between the DRM (magnetization of redeposited sediments) and SIRM: Rds = DRM/SIRM was used for the calibration (*Kurazhkovskii et al.*, 2011). The Rds is constant for marine gray-colored sediments redeposited in the same magnetic field (*Kurazhkovskii et al.*, 2011). When redeposition is performed in a magnetic field with the intensity of $50\mu T$, Rds = DRMt/SIRMt = 0.004. The temperature of the thermal cleaning (t) should not exceed $300°$ (at the heating to a higher temperature the probability magnetomineralogical changes rises). Since the sedimentary layer disclosed by borehole represented by the marine gray-colored aleurite, we used this ratio to calibrate paleointensity. In this case, the paleointensity was calculated using the equation: $H/H_0 = NRMt/DRMt = Rns/Rds = Rns/0.004$. The use of two calibration methods allowed for comparing of data paleointensity and assured their correctness.

**2.3.** For a detailed study of the behavior of the Early Paleogene paleointensity we used samples of sedimentary rocks from the Saratov region. According to *Sidorenko* (1967), the formation of this sedimentary layer occurred at the end of the Selandian. The sedimentary layer consisted of gray-colored aleurites with thickness of 10 m. The sampling was conducted at 30 cm intervals. The ores were sampled from each layer from which four to five samples were cut off for further study. Paleo - and petromagnetic study of samples of this sedimentary layer were performed in the Geophysical Observatory "Borok". The measurements of NRMt revealed that the sedimentation took place mainly during the normal polarity of the geomagnetic field (Fig. 2).

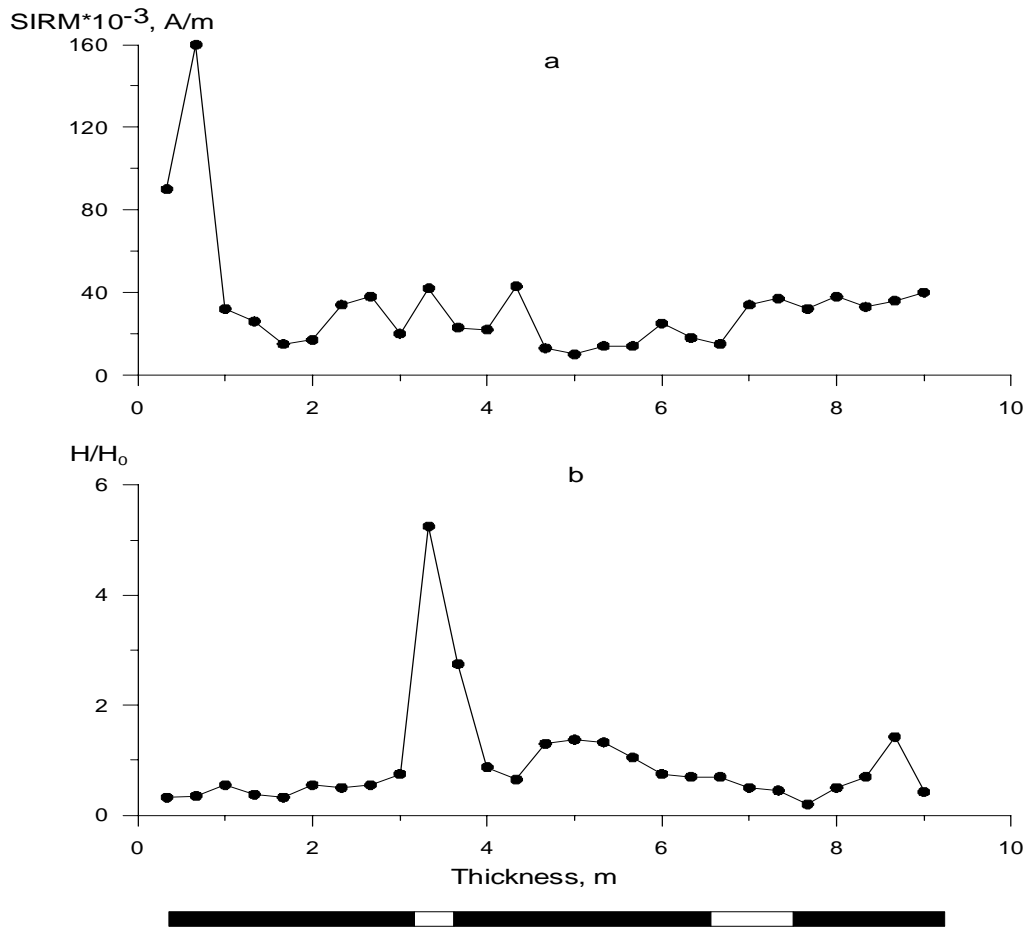

**Figure 2.** Variations in the SIRM (a) and paleointensity $H/H_0$ (b) by sediments of the Selandian sampled in the Saratov region. The polarity of the geomagnetic field is shown below the abscissa axis: normal polarity by black color, reverse polarity by white color.

It is likely that this interval can be identified with the C26n polarity chron. According to chronostratigraphic scale (*Zhamoida et al.*, 2000) at the end of Selandian the interval of the normal polarity (C26n) of the geomagnetic field with duration of about a million years was took place. Presumably, the time of sediments accumulation between two studied layers was about 30 thousand years.

Microprobe analysis (on the electron-probe analyzer Tescan Vega II) revealed that broad spectrum of allothigenic minerals of ilmenite and titanomagnetite lines served as carriers of the remanent magnetization of these deposits. In addition, in lower half of the sedimentary layer contained large amount of native iron. The source of origin of native iron is unknown, but it certainly could not be authigenous. The upper part of the sedimentary layer was enriched with detrital ilmenite. The set of detrital magnetic minerals was indicated the orientational nature of NRM. In addition, as at the study of borehole deposits, low inter-layer concentration of distribution of the vectors of magnetization directions and low values of the Königsberger factor also indicated the orientational nature of NRM.

The changes of the SIRM and $H/H_0$ parameters on thickness of sedimentary layer are given in Fig. 2. As can be seen from Fig. 2, the behavior of $H/H_0$ is not related to variations magnetomineralogical composition. Redeposition of these deposits laboratory following method by *Kurazhkovskii et al.* (2011) revealed that the ratio between DRMt and SIRMt is 0.004 (DRMt/SIRMt = 0.004). The paleointensity of these deposits as well as of the borehole deposits was calculated using the equation: $H/H_0 = Rns/0.004$.

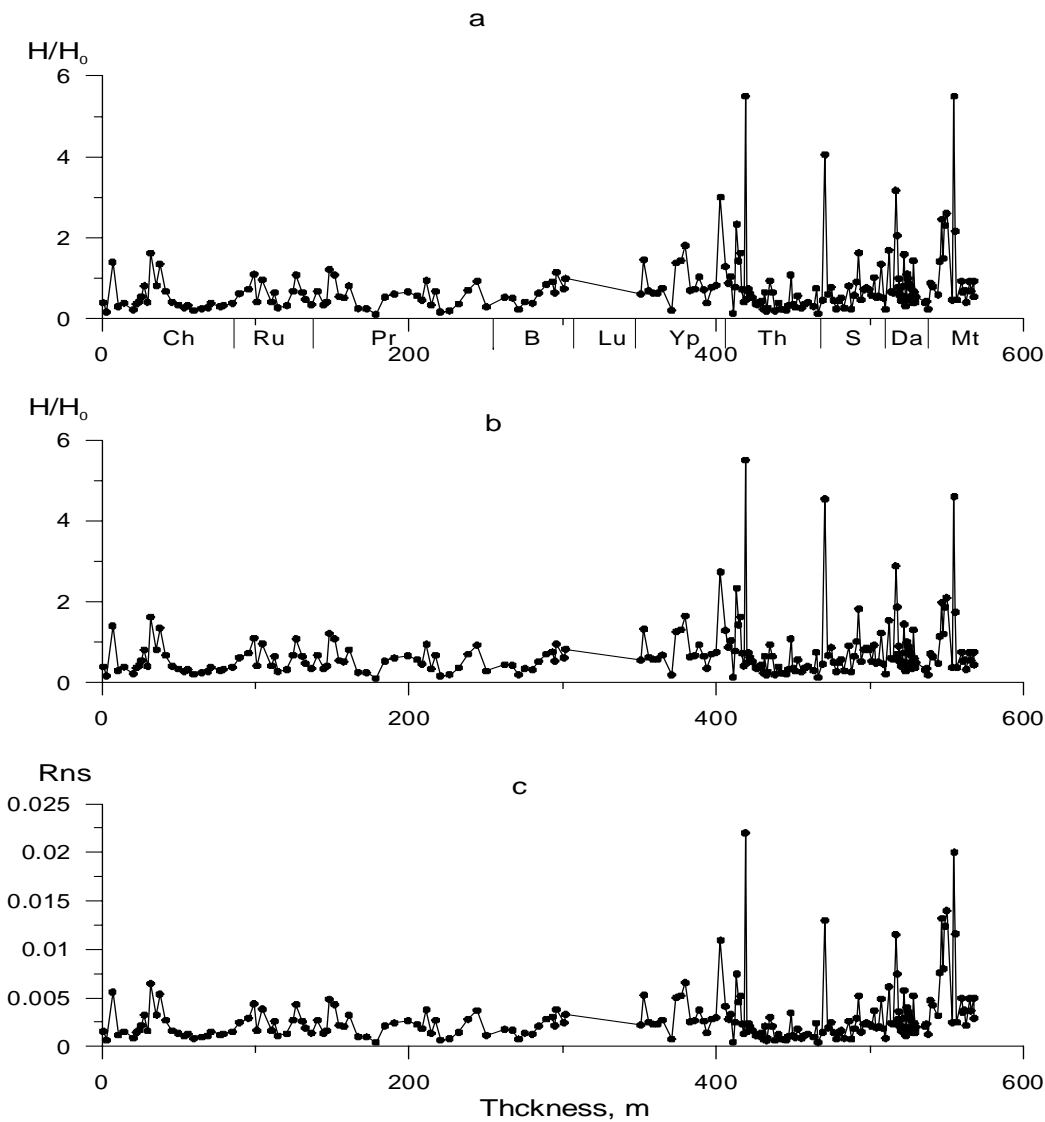

**Figure 3.** Behavior of the $H/H_0$ (a, b) and Rns (c) by sediments from borehole. a) Calibration of the paleointensity with using of Rds coefficient, b) calibration of the paleointensity with using DB PINT10.

## 3. Paleointensity behavior

The dynamics of the "absolute" values of the paleointensity obtained by various ways of a calibration and behavior of the Rns parameter given in Fig. 3 show that the behavior of paleointensity did not depend on the method of its calibration. At the same time, it is quite obvious that at the beginning and end of the Paleogene the behavior of the geomagnetic field was different, e. g. in the Paleocene (Danian - Thanetian) significant bursts of the paleointensity (up $5H_0$) of short duration took place. In the Eocene (Ypresian - Priabonian) maximal values of the paleointensity did not exceeded $2.5H_0$. At the end of the Paleogene the amplitude of paleointensity variations decreased. In the Miocene the maximal values of the paleointensity did not exceeded $1.5H_0$. Thus, the amplitude of paleointensity variations in the beginning of the Paleogene was significantly greater than in the end of the Paleogene.

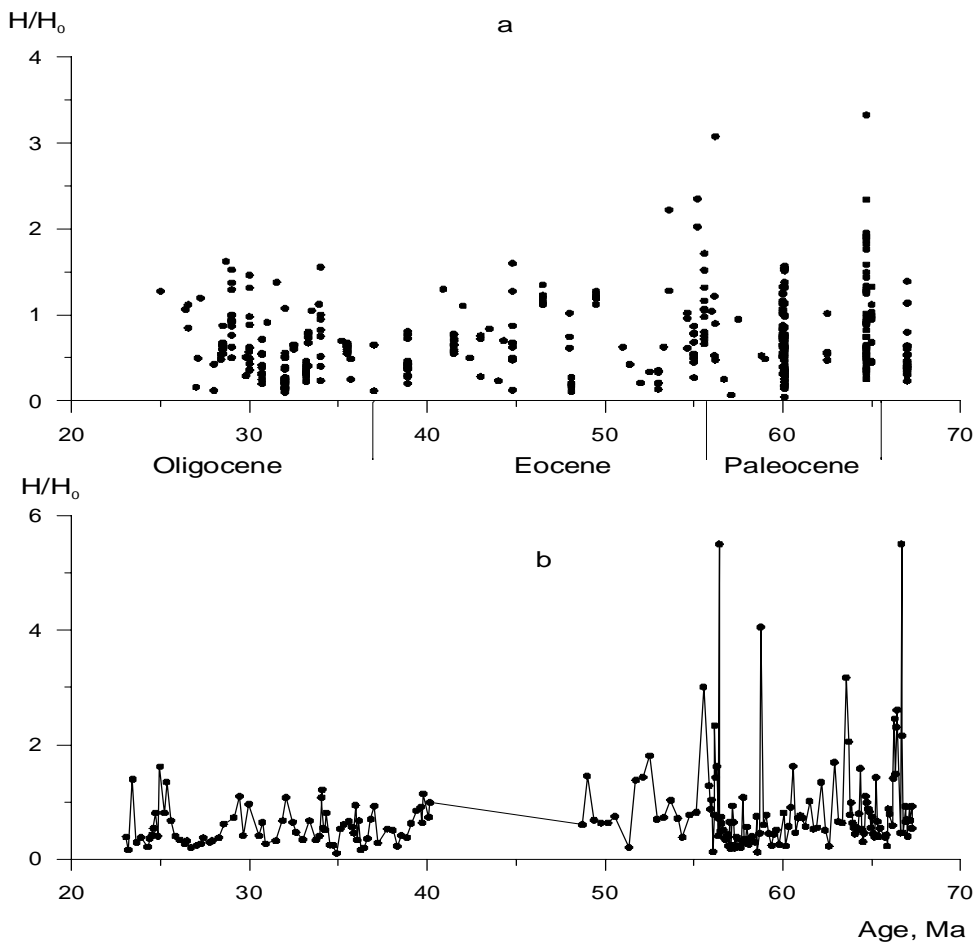

**Figure 4.** Comparison of the paleointensity data ($H/H_0$) obtained from (a) thermomagnetized rocks (PINT10) and (b) sedimentary rocks of the borehole.

The boundaries of series and stages have been isolated from the biostratigraphic studies of the borehole deposits. We brought them to conformity with the boundaries of geological epochs and ages, using chronostratigraphic scale (*Zhamoida et al.*, 2000). The values of the paleointensity from database PINT10 are shown in Fig. 4a; the changes of paleointensity (by the borehole) relative to the scale of geological age, in Fig. 4b. Comparison of the paleointensity data presented in Fig. 4a and Fig. 4b shows that the sedimentary and thermomagnetized rocks may lead to approximately the similar concepts concerning the changes occuring with characteristic times of the order of geological epochs. As it seen from Fig. 4 in the Paleocene paleointensity was on average higher than in the Oligocene. The paleointensity data obtained by thermomagnetized rocks are distributed on the scale of geological time irregularly. At that, it is possible to see that in the Paleocene bursts of paleointensity (by sediments) were followed by increase in the number and scattering of data obtained from the thermomagnetized rocks. This may indicate that the volcanic activity is associated with variations of the paleointensity.

Using the data obtained from sediments approximate characteristic times of variations of the paleointensity can be estimated. According to Fig. 4b characteristic times of variations of the paleointensity were a few millions years. The variations with approximately the same characteristic times have been obtained earlier at studying the behavior of paleointensity in the interval 33 - 22 Ma (*Constable et al.*, 1998).

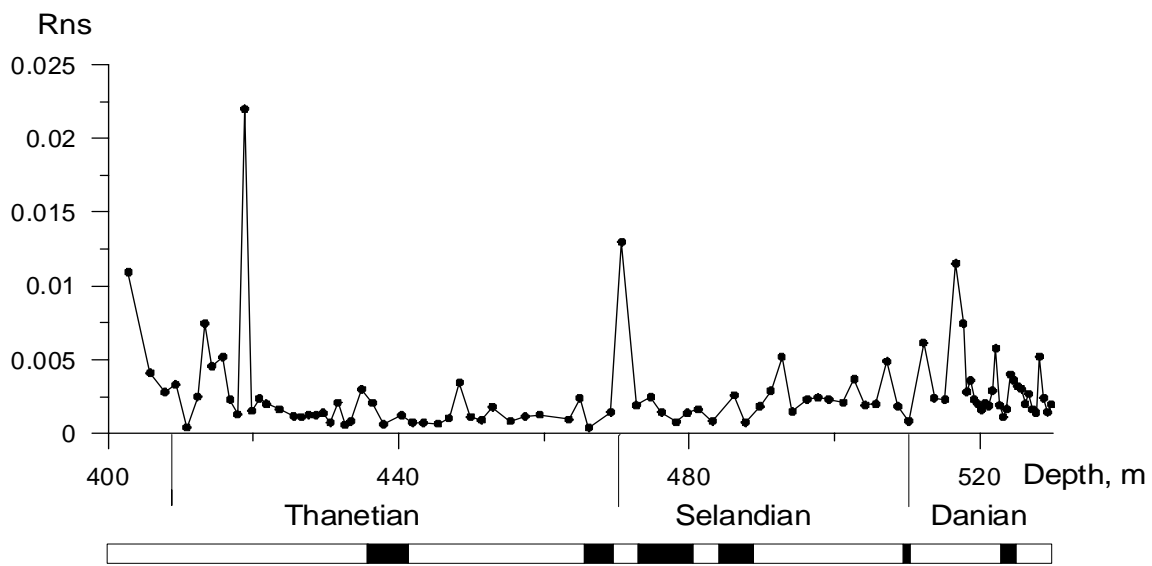

**Figure 5.** Behavior of the Rns parameter in the Paleocene from borehole sediments. The polarity of the geomagnetic field are shown below the abscissa axis: normal polarity by black color, reverse polarity by white color.

The available materials allowed comparing of data obtained from different sedimentary layers in the Selandian. In the paper (*Bogachkin*, 2004) it was revealed at the study of borehole sediments that at the end of Selandian age episode of normal polarity of the geomagnetic field took place. Accumulation of upper Selandian sediments from the Saratov region also occurred during the interval predominantly of normal polarity. In both cases these episodes of normal polarity have been identified with C26n chron. The detailed data of the behavior of the Rns parameter in the Paleocene, obtained from deposits of geomapping borehole are given in Fig. 5. Differences in the detail of the sampling (Fig. 2 and Fig. 5) do not allow for detailed comparison. However, both massifs of data indicate that at the end Selandian a significant burst in paleointensity took place.

## 4. Discussion

Mechanisms of formation of the remanent magnetization of sedimentary and thermomagnetized rocks have some fundamental differences. Consequently, the methods used for determination of the paleointensity by sedimentary and thermomagnetized rocks differ. The examples of methods for determining of the paleointensity by thermomagnetized and sedimentary rocks are given in many papers, e.g. in (*Perrin and Schnepp*, 2004; *Kurazhkovskii et al.*, 2011). In paleomagnetology the external convergence serves as main criterion for the correctness of the results: the results of a study of the same geomagnetic phenomena executed by different methods from different geological objects should coincide. We can not to compare exactly the results of determinations of the paleointensity by sedimentary and thermomagnetized rocks since the errors of the identification of age of geomagnetic events are usually larger than the characteristic times of the paleointensity variations. In our opinion, it is possible to compare the average (for the geological age) values of the paleointensity obtained by sedimentary and thermomagnetized rocks.

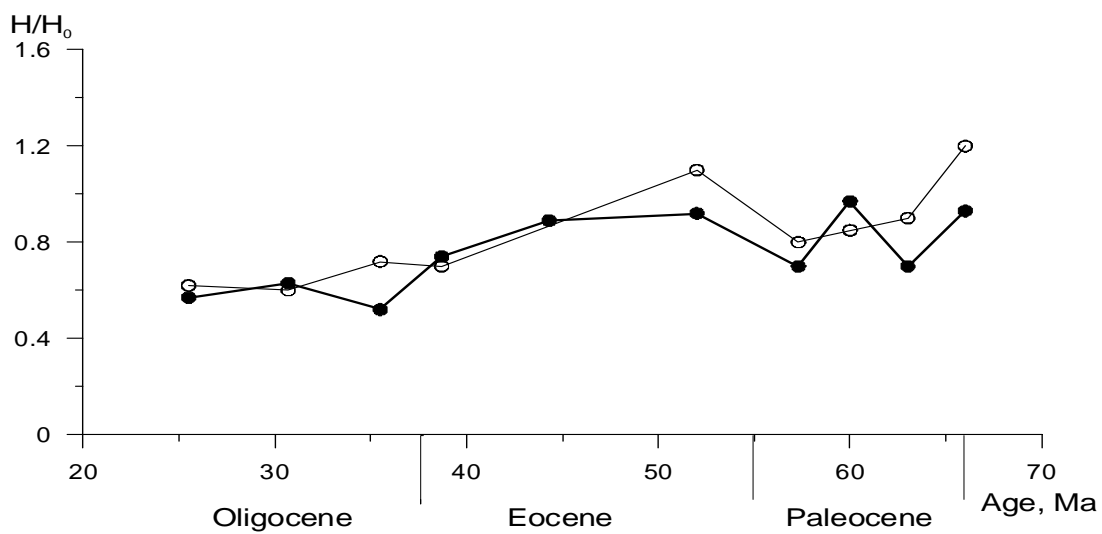

**Figure 6.** Mean values of paleointensity for the geological age by sedimentary (thin line) and thermomagnetized (thick line) rocks in the Paleogene.

Estimates of the average values of the paleointensity in the Paleogene from sedimentary (on data to Fig. 3a) and thermomagnetized (from PINT10) rocks are given in Fig. 6. As can be seen from Fig. 6, sedimentary and thermomagnetized rocks lead to the same conception about the main tendencies of changes of the paleointensity during the Paleogene: both massifs of data indicate a decrease in paleointensity during this period.

The presence of paleointensity bursts is the main feature of the behavior of paleointensity in the beginning of the Paleogene (in the Paleocene). Several reasons may substantiate the conclusion on their reality. The bursts of paleointensity do not coincide with any peculiarities of the behavior of petromagnetic parameters (Fig. 1). It is worth noting that a significant increase in K precedes the burst of paleointensity in the end of the Thanetian. At the same time, similar the changes of K in the Rupelian - Priabonian were not reflected on the peculiarities of the behavior of the paleointensity. Higher values of TK usually indicate high activity of authigenic processes. As follows from Fig. 1, high activity of authigenic processes (in the end of the Paleogene) did not lead to an increase in the Rns parameter value by which the behavior of the paleointensity was identified. Absence of coincidence of paleointensity bursts with variations of mineralogical composition of deposits may serve as an evidence of the reality of their existence.

The external convergence may serve as a test for the results of paleointensity determinations obtained on deposits of borehole by two time intervals. For instance, the lower part of investigated core was represented by sediments of the Late Maastrichtian (Fig. 3a and Fig. 4b). Earlier, we studied these deposits on the example of the Crimea deposits (*Kurazhkovskii et al.*, 2012). The study by described in *Kurazhkovskii et al.* (2012) as well as the present study it revealed that significant bursts of paleointensity took place in the end of the Maastrichtian. In addition, the burst of the paleointensity in the end of the Selandian was detected both on sediments of borehole and on the deposits of lower Volga (Fig. 2b, Fig. 3a and Fig. 5).

The study of the behavior of the Paleogene paleointensity during bursts can not be considers as completed. First, the sediments of borehole were studied in insufficient detail to claim that we have found the largest values of Paleogene paleointensity. Second, we are not sure that all the bursts of paleointensity which occurred in the Paleogene are detected. Third, the estimations of the values of the paleointensity were made based on the assumption that the NRM and H are related by linear dependence. There are cases when this relationship was not linear (*Tucker*, 1980; *Kurazhkovskii*, 1998). However, it is obvious that during bursts the paleointensity was several times higher than the average for the geological age.

High (about $3H_0$) paleointensity values were found in the Paleocene also its study by thermomagnetized rocks (Fig. 4a). The number of similar values of paleointensity determined by thermomagnetized rocks is still lower comparing to those obtained on the sedimentary rocks. It should be noted that the paleomagnetic data by thermomagnetized rocks (DB) are also much less abundant than the data which we obtained by studies of the borehole sediments.

Investigations of paleointensity by *Tauxe and Hartl* (1997) and *Constable et al*. (1998), as well as the results of the present study did not detect the paleointensity bursts at the end of the Paleogene. This favors the assumption that the behavior of the paleointensity at the beginning and end of the Paleogene were really different.

In our view, the studies of paleointensity by sedimentary rocks have a several advantages. For example, sedimentary rocks make it possible to estimate the amplitudes and the characteristic times of variations of the paleointensity in the Paleogene. It is not yet possible make similar estimates for thermomagnetized rocks. In addition, Fig. 4 b shows that sedimentary rocks at even a single sedimentary layer may lead to better idea about the changes of paleointensity in the Paleogene.

## 5. Conclusion

The patterns of the paleointensity behavior at the beginning and end of the Paleogene were different. In the beginning of the Paleogene (Paleocene) an alternation of paleointensity variations of small amplitude (about $0.5H_0$) with its bursts (up to $5H_0$) took place. In the end of the Paleogene bursts of the paleointensity were not detected. Its variations were occuring with the amplitude of the order of $0.5H_0$.

## Acknowledgments

We are grateful to the staff members of Saratov State University A. B. Bogachkin and A. Yu. Guzhikov for providing us with collections of samples of sedimentary rocks and for the materials of petromagnetic and stratigraphic studies. We would also like to acknowledge V.A. Tselmovich for study on the mineralogical composition of samples using electron-probe analyzer Tescan Vega II.